# HourGlass: Predictable Time-based Cache Coherence Protocol for Dual-Critical Multi-Core Systems

Nivedita Sritharan, Anirudh M. Kaushik, Mohamed Hassan, and Hiren Patel
{nivedita.sritharan, anirudh.m.kaushik, mohamed.hassan, hiren.patel}@uwaterloo.ca
University of Waterloo, Waterloo, Canada

***Abstract*—We present a hardware mechanism called HourGlass to predictably share data in a multi-core system where cores are explicitly designated as critical or non-critical. HourGlass is a time-based cache coherence protocol for dual-critical multi-core systems that ensures worst-case latency (WCL) bounds for memory requests originating from critical cores. Although HourGlass does not provide either WCL or bandwidth guarantees for memory requests from non-critical cores, it promotes the use of timers to improve its bandwidth utilization while still maintaining WCL bounds for critical cores. This encourages a trade-off between the WCL bounds for critical cores, and the improved memory bandwidth for non-critical cores via timer configurations. We evaluate HourGlass using gem5, and with multi-threaded benchmark suites including SPLASH-2, and synthetic workloads. Our results show that the WCL for critical cores with HourGlass is always within the analytical WCL bounds, and provides a tighter WCL bound on critical cores compared to the state-of-the-art real-time cache coherence protocol. Further, we show that HourGlass enables a trade-off between provable WCL bounds for critical cores, and improved bandwidth utilization for non-critical cores. The average-case performance of HourGlass is comparable to the state-of-the-art real-time cache coherence protocol, and suffers a slowdown of 1.43× and 1.46× compared to the conventional MSI and MESI protocols.**

## I. Introduction

Modern mixed-criticality systems (MCS) consist of a combination of critical and non-critical tasks that share the hardware resources when deployed on a computing platform. For certification, it is imperative to guarantee that critical tasks never exceed their requirements. Otherwise, the system may result in catastrophic consequences. Example domains include avionics and automotive (DO 178C and ISO 26262). Non-critical tasks, however, do not require such stringent guarantees, but instead, may only require best-effort service such as improved average-case performance. A primary concern in MCS is the resulting interference caused by the critical and non-critical tasks sharing the hardware resources. This is because the non-critical tasks can contribute to the worst-case execution time (WCET) of the critical tasks or it may significantly complicate the WCET analysis. The contributing interference may originate from one or many of the hardware components such as the main memory, last-level caches, and shared buses. Consequently, there is considerable interest in devising strategies to mitigate the impact of such interference.

As demands for more functionality from MCS increase, there is noticeable interest in leveraging multi-core platforms to deploy such applications (Freescale P4080, [1]). While using multi-cores reduces hardware cost, it also offers parallelism that applications can exploit. However, they also introduce additional sources of interference. One such interference occurs when multiple tasks deployed on different cores access shared data that is temporarily stored in the private cache of a core. Since multiple threads may update the shared data, ensuring that any read of the shared data receives the most up-to-date write to the shared data is essential for correct execution. This involves transferring the data from one private cache, through the bus, to the requesting core's cache. Since the worst-case latency (WCL) to access data is a component of the WCET, predictably managing transfers of shared data between critical and non-critical tasks remains a challenging problem [2].

Recently, a hardware-based cache coherence protocol to predictably manage accesses to shared data in multi-cores called PMSI [3] was proposed. However, PMSI assumed only critical tasks, and a non-critical task would be treated as a critical task. As a result, a non-critical task would cost the critical tasks an opportunity for tighter worst-case latency bounds. Furthermore, PMSI does not offer any mechanisms to encourage non-critical tasks to improve its average-case memory bandwidth utilization.

In this work, we propose HourGlass, a predictable cache coherence protocol that is criticality-aware. Specifically, we design HourGlass to be a *dual-critical multi-core system* (DCMS) that operates on only two criticality levels: critical (cr) and non-critical (ncr). For tasks on the critical cores, we ensure worst-case latency bounds. For tasks on the non-critical cores, we provide no guarantees. However, we equip HourGlass with a timer-based mechanism to allow for tasks on non-critical cores to improve their memory bandwidth utilization. The timers allow for a trade-off between potentially improving memory bandwidth utilization of non-critical tasks while loosening the WCL bounds of critical tasks. This is acceptable if the critical tasks continue to meet their temporal requirements. HourGlass also allows for predictable sharing of data between critical tasks, and between non-critical and critical. We provide a timing analysis of HourGlass, and describe the various latency components that contribute to the WCL of requests from critical tasks. We prototype HourGlass in gem5 [4], a cycle accurate micro-architectural simulator, and evaluate with multi-threaded benchmarks from SPLASH-2 [5], and synthetic benchmarks that exhibit maximum data sharing. We observe that the WCL of all memory requests from critical tasks are within their analytical bounds, and that timers do encourage a trade-off between bandwidth utilization of requests from non-critical cores and WCL for critical ones.

## II. Related Work

Multi-core real-time platforms contain interconnects, caches, and main-memory that are shared by the cores in the platform. These components are common sources of interference when multiple cores make accesses to the shared memory resulting in non-trivial timing analysis of shared data accesses. Authors in [6], [7] propose disabling shared data in the private cache hierarchies to provide timing analysis for memory requests to shared data at the cost of long execution times. Alternatively, solutions proposed in [8], [9], [10] make operating system (OS) changes to include data-sharing aware scheduler policies that identify tasks that share data, and avoid running these tasks concurrently. However, these solutions require changes to the OS, and hardware support from performance counters in order to identify tasks that share data. Another approach proposed in [11] modified the application such that accesses to shared data were protected using lock mechanisms such that the shared data was accessed by only one core at any time instance. This solution was a software implementation of cache coherence. In the worst-case, this approach performs as well as sequential execution of tasks sharing data.

A recent solution by Hassan et al. [3] proposed a hardware cache coherence protocol called PMSI to manage shared data accesses. PMSI required hardware modifications, but no changes to the OS and application. Compared to prior approaches, PMSI allowed tasks to simultaneously access copies of shared data cached in their private caches resulting in improved average-case execution time. However, PMSI was not designed for mixed-criticality systems. We believe that supporting criticality awareness in the cache coherence protocol is essential for modern real-time systems [2]. In this work, we present HourGlass, a cache coherence protocol that is criticality-aware with support for two criticality levels, which offers tighter worst-case latencies for critical cores.

There are several research efforts that provide explicit support for different criticalities in bus arbiters [12], [13], [14], [15]. Paolieri et al. [12] presented a dual-criticality arbitration where hard real-time cores (HRT) used round-robin arbitration, and whenever there were no requests from HRT, non-critical cores are serviced yielding a worst-case bound for HRT. Hassan et al. [13] presented CArb, a statically scheduled two-level bus arbitration scheme that used harmonic weighted round-robin (HWRR) arbitration. The first level performed HWRR amongst criticality classes, and the second between tasks within a criticality class. Cilku et al. [14], [15] proposed a two-layer arbiter that distinguished between memory requests from critical and non-critical cores. Their approach used TDM arbitration such that critical cores were pre-assigned slots in the TDM schedule after which a fixed number of slots were reserved to service requests from non-critical cores. Within the slots reserved for non-critical cores, round-robin was performed. HourGlass takes inspiration from Paolieri et al. [12], and Cilku et al. [14], [15], but we require no pre-allocated slots for ncr cores, which offers further opportunities for tighter worst-case latency bounds for the critical cores.

## III. Background

### A. Cache coherence

Multi-core systems share data between cores by accessing addresses within a shared address space. Modern multi-core platforms implement private cache hierarchies that exploit spatial and temporal locality to improve the application's performance. Hence, shared or private data may reside in the private cache hierarchy of multiple cores. This allows multiple copies of the data to exist in the private cache hierarchies of different cores. For correct execution of parallel programs with shared data, it is essential that any core performing a read operation on the shared data obtains the most recent write to the shared data. Since multiple copies of the shared data are privately cached across multiple cores, there must be a mechanism to ensure that reads to the shared data receive the most up-to-date data. This is known as keeping data values *coherent across multiple cores*.

A mechanism known as *cache coherence* is a solution that keeps shared data values coherent across multiple cores [16]. There are software and hardware techniques to implement cache coherence. Software approaches for cache coherence require application changes to manage the various copies of shared data explicitly. On the other hand, a hardware implementation of cache coherence implements a protocol that enforces strict rules governing the coherent view of multiple cached copies of data across multiple cores. A hardware cache controller implements this protocol. Hardware cache coherence works at the granularity of cache lines, which is the unit of data transfer in the memory hierarchy. Based on the implementation, hardware cache coherence can be realized using a snoopy shared bus, which allows all cores to observe memory activity of other cores, or using a centralized directory that co-ordinates the coherence across multiple cores [16]. In this work, we focus on hardware cache coherence that use a snoopy shared bus implementation, which we find is appropriate for current multi-core platforms with eight cores or less (Freescale P4080, [1]) used in real-time multi-core systems.

### B. Coherence protocol

A cache coherence protocol is an implementation of a set of rules that ensure data coherence. This set of rules identifies states that denote the read and write permissions of cache lines, and transitions between these states that occur due to activities of other cores in the system on the same cache line. The **M**odified-**S**hared-**I**nvalid (MSI) cache coherence protocol is a fundamental cache coherence protocol that several modern cache coherence protocols are based upon such as the MESIF, and MOESI protocols [16]. MSI consists of three *stable states*. The semantics for each of these states is as follows: 1) *Invalid* ($I$) indicates that the cache line does not have valid data. 2) *Modified* ($M$) represents that the core has modified the cache line data; hence, it has the most up-to-date data. Only one core can have a cache line in the modified state. 3) *Shared* ($S$) identifies that the cache line was read, but not modified. Multiple cores may have the same cache line in the shared state. This allows read hits in their respective private caches. Cache coherence protocol also consists of *transient states*, which are intermediate states between stable states. These states represent whether the core is waiting for a data response, or waiting for the memory requests to be ordered on the shared bus. For example, the transient state $IM^D$ is an intermediate state between the invalid ($I$) state and modified ($M$) state. It denotes that a core observed its write request to a cache line on the bus, and awaits a data response. A cache line changes states based on the activities of other cores. Transitions between states occur by exchanging coherence messages between the cores and shared memory.

When discussing coherence activities, we refer to the *private core* to identify the core whose cache controller is under consideration, and *remote core*s as all other cores. Since every private cache implements the same set of rules in its cache controller, we distinguish requests made by the private core and those by the remote cores. To accomplish this, a core views coherence messages on the bus as either *Own* or *Other* coherence messages. *Own* denotes that the cache controller observes a coherence message generated by its private core, and *Other* as a coherence message generated by remote cores.

## IV. System Model

We assume a cache-coherent multi-core real-time system with $N$ in-order cores $\{c_0, c_1, ..., c_{N-1}\}$. Each core has its own private instruction cache (L1-I) and data cache (L1-D). A core can only have one pending request to the bus. The shared memory is assumed to be the main memory. The cores and the shared memory are interconnected with a snooping bus, which exchanges coherence messages between them. Additionally, there is a common cache data bus connecting the cores to support cache-to-cache transfers. We assume that once a data transfer over the bus starts, it cannot be preempted. We characterize each core by its criticality level ($cl$) and the core identifier, $c_i^{cl}$ where $cl \in \{\mathsf{cr}, \mathsf{ncr}\}$ and $i$ indicates the core identifier. Memory request from cr cores must have WCL guarantees per request whereas memory requests from ncr cores have no guarantees. However, ncr can benefit from improved memory bandwidth. We require that a single task is mapped to a single core for the duration of the application's execution. However, the tasks distributed across the multiple cores can share data. We denote the number of cr cores as $N_{\mathsf{cr}}$, and ncr cores as $N_{\mathsf{ncr}}$, such that $N = N_{\mathsf{cr}} + N_{\mathsf{ncr}}$. Note that we use the term cr or ncr cores interchangeably with memory requests from cr or ncr cores.

Timers improve bandwidth utilization of ncr cores. Hence, each core contains timers that enable holding onto the cache line for a fixed time duration (cycles) even when it is requested by a remote core. Once the timers expire, the private core responds to the coherence messages from the requesting remote cores. We associate two timers with each cache line for a core. The countdown timers are denoted as $t_n(c_n^{cl}, c_m^{cl})$ where $t_n$ identifies the current value of the timers for $c_n$ where $n, m \in \{0, 1, ..., N-1\}$. $c_n$ denotes the core that has a valid copy of the shared data, and $c_m$ denotes the requesting core. For example, assume that $c_i$ is a cr core, and $c_j$ is a ncr core. Then, $t_i(c_i^{\mathsf{cr}}, c_j^{\mathsf{ncr}})$ denotes the timer configuration for $c_i$ when a cr core $c_i^{\mathsf{cr}}$ has a valid cache line, and observes a memory request from a ncr core $c_j^{\mathsf{ncr}}$. The timers are initialized with timeout values on receiving a valid copy of the cache line. We denote $v(\mathsf{cr}, \mathsf{cr})$ as the initial timeout value for a cache line present in the private cache of a cr core, and requested by another cr core, and $v(\mathsf{cr}, \mathsf{ncr})$ as the initial timeout value for a cache line present in the private cache of a cr core, and requested by a ncr core. Similarly we have $v(\mathsf{ncr}, \mathsf{cr})$ and $v(\mathsf{ncr}, \mathsf{ncr})$ for cache lines present in the private cache of a ncr core, and requested by another cr core and ncr core, respectively. The timeout values are precomputed based on the properties of the DCMS. Since we assume two criticality levels, a valid cache line in a cr core $c_i^{\mathsf{cr}}$ may receive requests for the cache line from a core of either of the criticality levels resulting in two timer configurations $t_i(c_i^{\mathsf{cr}}, c_j^{\mathsf{cr}})$, and $t_i(c_i^{\mathsf{cr}}, c_k^{\mathsf{ncr}})$ where $c_k$ is another ncr core. Notice that these two timer configurations identify the requesting core to be either cr and ncr respectively. Similarly, a request to a valid cache line in a ncr core $c_j^{\mathsf{ncr}}$ also has two timer configurations $t_j(c_j^{\mathsf{ncr}}, c_i^{\mathsf{cr}})$, and $t_j(c_j^{\mathsf{ncr}}, c_l^{\mathsf{ncr}})$ where $c_l$ is another ncr core. We assume that the timer configurations are set based on the requirements of the application, and they are not changed during application execution. Moreover, the timers are private to the core; hence, their values are not communicated to other cores.

The shared snooping bus uses time-division-multiplexing (TDM) to arbitrate accesses to the shared memory between cores. cr cores are pre-assigned one TDM slot while ncr cores are granted access to the shared bus during TDM slots that have no pending request from cr cores, which we call *slack slots*. The TDM bus arbitrates memory requests, data responses, and coherence messages. We assume the TDM slot-width $SW$ to be large enough to complete one data transfer from the shared memory to the private cache of a core, and the transfer of any necessary coherence messages. The transfer of coherence messages and data responses begin at the start of a TDM slot. We also assume that a core can issue only one memory request in its designated TDM slot.

## V. Motivating Examples

Cache coherence maintains a coherent view of multiple copies of shared data by systematically propagating changes of one copy to the other copies. In conventional cache coherence protocols [16], the propagation of changes occurs independent of the criticality of the core that either has a valid copy of the data or requests it. Therefore, current coherence techniques do not provide a mechanism to distinguish accesses from cr

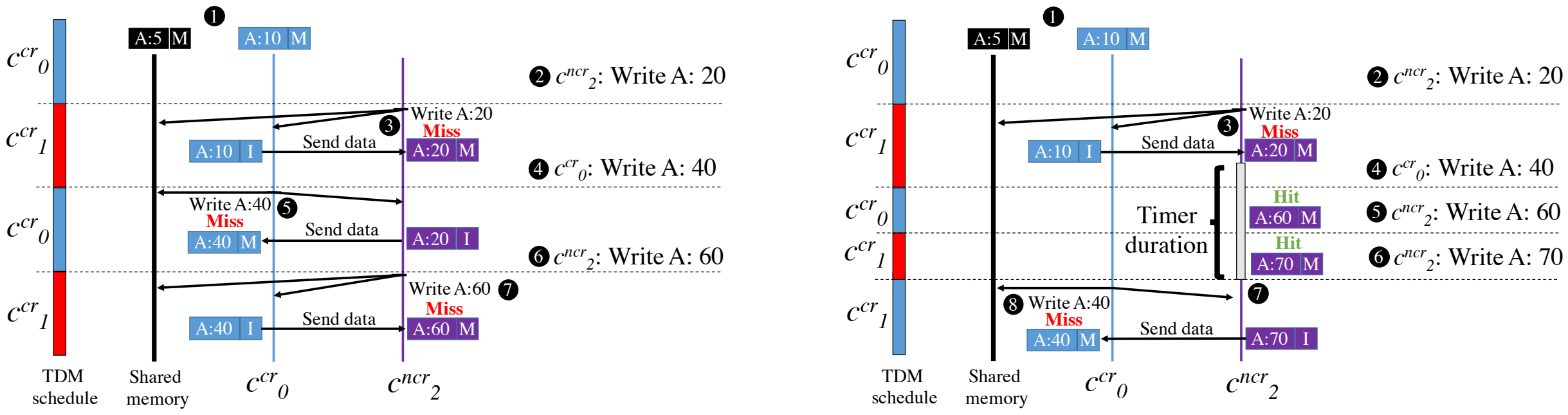

(a) PMSI cache coherence protocol with fixed priority arbitration.

(b) Proposed solution with timers.

Fig. 1: Supporting criticality awareness in cache coherence protocols.

cores over ncr cores. Even the state-of-the-art cache coherence protocol in real-time systems called PMSI [3] assumes all cr cores. The ability to identify requests from cr or ncr cores is essential in DCMSs. This is because a key principle in DCMS is to reduce the interference suffered by cr cores originating from ncr cores to assist in tighter worst-case bounds while encouraging ncr cores to meet its bandwidth requirements.

PMSI [3] does not differentiate criticality of cores. Therefore, a ncr core in PMSI would be treated as a cr core. This means the coherence state of a shared cache line of a cr core is affected by the activities of the ncr core by the mere fact that a TDM slot must be assigned to the ncr core. Consequently, we find that introducing criticality-awareness in cache coherence can provide tighter worst-case bounds for cr cores. An alternative is to make modifications to PMSI to support different criticality levels. Suppose that we change the bus arbitration policy to operate as presented in Section IV. The bus arbitration presented is criticality-aware. In contrast to TDM as used in PMSI, this arbitration does not require pre-allocated slots to ncr cores as their requests are only serviced when the cr cores do not generate memory requests.

Using Figure 1a, we illustrate that this implementation allows for tighter worst-case latency (WCL) bounds for cr cores. For the sake of simplicity, we assume that PMSI also supports cache-to-cache transfers between cores, which it originally does not as presented by Hassan et al. [3]. We assume a dual-critical multi-core system of cr cores $c_0^{\mathsf{cr}}$, and $c_1^{\mathsf{cr}}$, and ncr cores $c_2^{\mathsf{ncr}}$, and $c_3^{\mathsf{ncr}}$. We only show the core activity for cr core $c_0^{cr}$ and ncr core $c_2^{ncr}$. Initially, $c_0^{\mathsf{cr}}$ has data $A$ in modified state ❶. At ❷, $c_2^{\mathsf{ncr}}$ has a write request. Since, $c_1^{\mathsf{cr}}$ does not have a pending memory request, the arbiter uses the slack slot to issue $c_2^{\mathsf{ncr}}$'s write request to $A$ ❸. Since $c_0^{\mathsf{cr}}$ has the most up-to-date copy of $A$, it sends the data to $c_2^{\mathsf{ncr}}$, and invalidates its copy ❸. At ❹, $c_0^{\mathsf{cr}}$ has a write request to $A$. At ❺, $c_0^{\mathsf{cr}}$'s write request is placed on the shared bus, and $c_2^{\mathsf{ncr}}$ sends the up-to-date data to $c_0^{\mathsf{cr}}$, and invalidates its copy. At ❻, $c_2^{\mathsf{ncr}}$ has a write request to $A$. Since, $c_1^{\mathsf{cr}}$ does not have a pending request, the bus arbitration grants this slack slot to $c_2^{\mathsf{ncr}}$'s pending write request, and completes its write request at ❼. Note that this example performs coherence correctly, but it requires a minimum of two TDM slots to be assigned to the cr cores. With PMSI, all four cores would need to be assigned at least one slot. Although this is a simple example, it illustrates that we can obtain tighter WCL bounds on cr cores.

Note that we only ensure WCL bounds for cr cores, and no guarantees on either WCL or bandwidth for the ncr cores. However, we do wish to provide ncr cores a mechanism to improve its memory bandwidth utilization while still ensuring WCL bounds for cr cores. For example, in Figure 1a, consider that the first write request to $A$ from $c_2^{\mathsf{ncr}}$ (❷) is succeeded by multiple read and write requests to $A$ from $c_2^{\mathsf{ncr}}$ (not shown). Currently, requests from cr cores to the same data results in $c_2^{\mathsf{ncr}}$'s copy of $A$ to be invalidated. This in turn results in cache misses for succeeding requests to $A$ for the $c_2^{\mathsf{ncr}}$. A mechanism that we investigate in this work is to allow cores to hold shared data in their private caches for a pre-defined time duration. The time duration still guarantees WCLs for cr cores, but it allows ncr cores to leverage cache hits for repeated data accesses resulting in increased memory bandwidth utilization.

Figure 1b shows the use of timers with the proposed criticality-aware bus arbitration policy. We assume that only the ncr cores have timers. The initial states (❶, ❷) are similar to the example in Figure 1a. At ❸, $c_2^{\mathsf{ncr}}$'s write request is satisfied in the slack slot of $c_1^{\mathsf{cr}}$. The difference from the previous example is that $c_2^{\mathsf{ncr}}$ holds $A$ for a time duration of 1 TDM period. In other words, at the end of the next slot of $c_1^{\mathsf{cr}}$, $c_2^{\mathsf{ncr}}$ self-invalidates its copy of $A$ in its private cache. During this time period, successive write requests from $c_2^{\mathsf{ncr}}$ to $A$ are cache hits ❺,❻ resulting in high memory bandwidth utilization. During this timer period, the response for the write request from $c_0^{\mathsf{cr}}$ ❹ is deferred until the end of the timer duration or timeout ❼. At the end of the timer duration, $c_2^{\mathsf{ncr}}$ invalidates its copy of $A$ (❽), and sends it to the first pending cr core, which is $c_0^{\mathsf{cr}}$. Note that this reordering of memory requests between the ncr and cr cores is a valid reordering under sequential consistency [16]. Hence, timers introduce a trade-off between the WCL of memory requests from cr cores, and memory bandwidth utilization of ncr cores.

| Criticality | State | Bus events | | | |
|---|---|---|---|---|---|
| | | OtherGetS-cr | OtherGetS-ncr | OtherGetM-cr | OtherGetM-ncr |
| cr | $IS^D$ | - | - | update cr-Dest/$IS^DI$ | update ncr-Dest/$IS^DI$ |
| cr | $IS^DI$ | - | - | update cr-Dest | update ncr-Dest |
| cr | $IM^D$ | update cr-Dest/$IM^DI$ | update ncr-Dest/$IM^DI$ | update cr-Dest/$IM^DI$ | update ncr-DDest/$IM^DI$ |
| cr | $IM^DI$ | update cr-Dest | update ncr-Dest | update cr-Dest | update ncr-Dest |
| ncr | IS$^D$ | - | - | reissue GetS/$IS^{AD}$ | update cr-Dest/$IS^DI$ |
| ncr | $IS^DI$ | - | - | reissue GetS/$IS^{AD}$ | update cr-Dest/$IS^DI$ |
| ncr | $IM^D$ | reissue GetM /$IM^{AD}$ | update ncr-Dest/$IM^DI$ | reissue GetM /$IM^{AD}$ | update cr-Dest $IM^DI$ |
| ncr | $IM^DI$ | reissue GetM /$IM^{AD}$ | update ncr-Dest/$IM^DI$ | reissue GetM /$IM^{AD}$ | update cr-Dest $IM^DI$ |

TABLE I: Different HourGlass transitions based on core criticality.

## VI. HourGlass: A Time-based Predictable Cache Coherence Protocol

HourGlass is a predictable time-based cache coherence protocol for DCMSs. We derive HourGlass from the conventional MSI protocol. We design HourGlass such that there is a per request worst-case latency bound for crcores. For ncr cores, there are no guaranteed bounds, but we provide means to support improved memory bandwidth. For predictability, we must satisfy the invariants described in PMSI [3] for crcores, but we must also prioritize crcores over ncr cores.

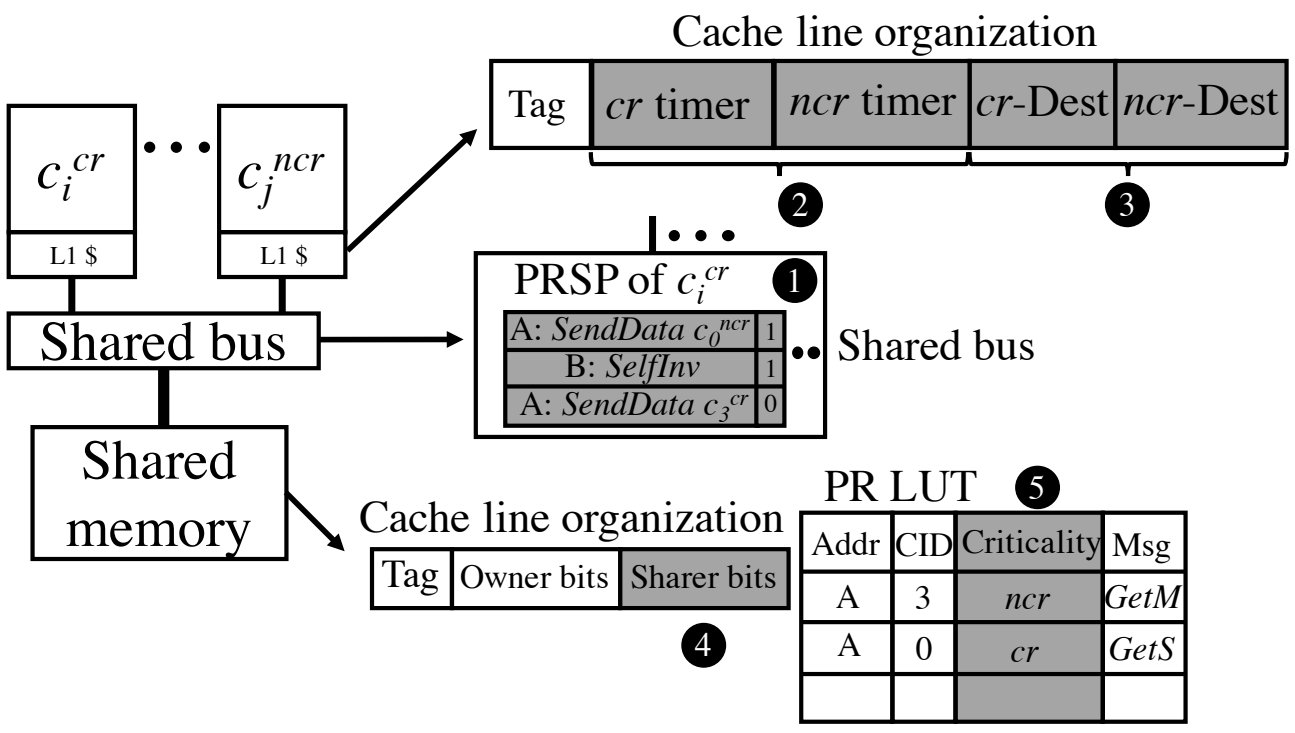

Fig. 2: Architectural modifications for HourGlass.

### A. Architectural Modifications

Figure 2 shows the architectural support necessary for HourGlass. The architectural changes can be categorized into modifications to the shared bus (Section VI-A1), and modifications to the cache controllers (Section VI-A2).

*1) Architectural modifications to shared bus:* Recall that we assume (Section IV) a criticality-aware arbitration scheme with cr and ncr cores. The arbitration assigns slots to only crcores, and services ncr cores in slack slots. Hence, we assume the hardware support to arbitrate across crcores, and service requests during slack slots or during assigned slots is already available in the shared bus. Since HourGlass is criticality-aware, this requires support to identify and reorder core destinations for cache-to-cache transfers and self-invalidation messages based on the criticality of the requesting core. For example, consider $c_0^{\text{cr}}$ has a modified copy of a cache line in its private cache. Further, consider two pending requests to the same cache line from $c_2^{\text{ncr}}$, and $c_1^{\text{cr}}$ that are observed in the same order by $c_0^{\text{cr}}$. We require HourGlass to identify and reorder data responses such that $c_0^{\text{cr}}$ sends the data to $c_1^{\text{cr}}$ first, and cancels the pending data response to $c_2^{\text{ncr}}$ although it observed the request from $c_2^{\text{ncr}}$ first. To support this, we add a pending response (PRSP) buffer for each core in the shared bus that holds coherence messages related to sending data to another core and to the shared memory, and self-invalidation messages (explained in Section VI-B) from cores after their timers expire. This is shown as ❶ in Figure 2. Each PRSP buffer is of size $N$. The arbiter identifies canceled pending data responses from ncr cores in the presence of data responses to crcores. This is done using a valid bit in the PRSP entries shown in ❶.

Adding PRSP buffers requires changes to the criticality-aware arbitration logic. Cache-to-cache transfers, write-backs to shared memory, and self-invalidations due to coherence in HourGlass are carried in the TDM slot of the remote core requesting the data or causing the self-invalidation. Hence, at the start of a TDM slot for a particular core $c_k$, the arbitration logic must index the PRSP buffers of all the cores to identify if there are pending cache-to-cache data responses with destination core as $c_k$ or self-invalidation responses for $c_k$. If there exists one, the data response to $c_k$ or self-invalidation is carried out in the TDM slot for $c_k$. If $c_k$ does not have a pending request, and no data responses or self-invalidation messages for $c_k$ exist in the PRSP buffers of other cores, this TDM slot is assigned to a ncr core (slack slot) in round-robin. The arbiter determines which ncr core to grant access to the bus at the beginning of the slack slot, and only ncr core may proceed. This ensures that the ncr core can complete within the slot.

*2) Architectural changes to cache controllers:* HourGlass uses timers to guarantee WCL bounds for crcores while simultaneously allowing ncr cores to receive improved memory bandwidth. These timers are typically available in most current commercial-off-the-shelf (COTS) micro-architectures in the form of time-stamp registers. For example, x86 architectures have the timer-stamp registers [17] that increments based on the core frequency, and most ARM architectures have a system counter that counts at a specific frequency [18]. We employ such high precision timers for HourGlass.

In order for cores to hold cache lines for a time duration, we extend the tag bits of cache lines to hold two timer values based on their criticality. For example, cache lines in crcore will have two countdown timers $t_i(c_i^{\text{cr}}, c_j^{\text{ncr}})$ (ncr timer) and $t_i(c_i^{\text{cr}}, c_k^{\text{cr}})$ (cr timer) where $i, j, k \in \{0, ..., N-1\}$. The initial values of the countdown timers are computed based on the TDM slot width. We assume that the counter values are represented as 64 bit values. This architectural modification is shown in Figure 2 as ➋. When a core receives a cache line from either the shared memory or from a remote core, the timer values are set with the timeout configured values, and begin to count down every cycle until they are zero.

We augment the cache controller with logic to identify the criticality of remote requests. We assume that cores are aware of the criticality of other cores present in the multi-core platform. This information can be configured at boot time, and stored in a dedicated on-chip read-only memory (ROM). The criticality of the requesting core is identified through the coherence messages (explained in Section VI-B). This allows a core that holds a valid copy of a cache line to differentiate the criticality of pending memory requests. The ability to differentiate criticality of pending memory requests is necessary for cache-to-cache transfers. This is because in the event of multiple pending requests for different critical cores to a cache line, the core that sends the cache line must send the data to the oldest pending crcore. To this end, we further extend the tag bits of a cache line to include two destination fields to denote the cr- and ncr cores requesting the cache lines as shown in Figure 2 as ➌. Hence, if both destination fields for a cache line have valid core identifiers, the cache line will be transfered to the crcore. Note that once a destination field is populated with a valid core identifier, it is not overwritten with another core identifier. This is done in order to service multiple pending requests to a cache line from the same criticality cores in a predictable manner.

We also track accurate information about the sharers of cache lines in the system at the shared memory (more details in Section VI-B). This is not necessary in conventional cache coherence protocols, and PMSI. Hence, the cache lines of the shared memory are extended by sharer bits that track the sharers of a cache line. This is shown as ➍ in Figure 2. The maximum number of sharer bits is equal to the number of cores in the multi-core platform. This is because multiple cores can have a cache line in the shared state ($S$), and timeout at different instances as each core may have received a copy of the cache line at different time instances. Knowledge about when no cores in the system have a valid copy of the data is important to HourGlass in order to avoid data incoherency. We explain this in detail in Section VI-B. HourGlass requires modifications to the pending request lookup table (PR LUT) in the shared memory. The PR LUT is an implementation of an invariant listed by Hassan et al. [3] for building predictable cache coherence protocols. The PR LUT manages multiple pending requests to the same cache line in a predictable manner. With support for criticality, we require modifications to the PR LUT such that pending requests from crcores to a cache line are serviced before ncr cores to the same cache line. Hence, we extend the PR LUT to include the criticality of the pending requests to a cache line shown as ➎ in Figure 2, and add logic to cancel pending ncr requests to a cache line on encountering a request to the same cache line by a crcore. Section A-D describes the hardware overhead of HourGlass for a four core system.

### *B. Coherence Protocol Modifications*

HourGlass requires modification to the MSI cache coherence protocol in addition to the architectural changes listed earlier for criticality awareness. This is because coherence messages and responses need to be differentiated based on the criticality of the cores. Tables I and A1 describe the different coherence states and transitions between states for the private caches. Core events denote activities of the core such as loads, stores, and replacements, and bus events denote coherence messages and data responses observed on the bus. We categorize the changes to MSI into two categories: changes for criticality awareness, and changes for timer support.

*1) Changes for criticality-awareness:* Recall from Section VI-A that the cache controllers have hardware support to differentiate memory requests from different criticality cores. This differentiation is carried out through the coherence messages sent by remote cores. Since each core is aware of the criticality of the other cores in the system, the core identifier information in coherence messages from remote cores is used to identify their criticality. Hence, we introduce two classes of coherence messages based on the criticality of the remote core denoted as $Message$-cr, and $Message$-ncr. For example, $OtherGetM$-cr denotes that a core with a valid copy of a cache line sees a store request from a crcore.

The shared memory identifies cache lines that are not cached in the private caches of cores ($I$ state), and when there exists at least one core with a valid read-only copy ($S$). In the conventional MSI cache coherence protocol and PMSI, the shared memory fuses the $I$ and $S$ states into one $I/S$ state. This is necessary in order to track multiple sharers of a cache line that can timeout at different time instances. The shared memory also has information about the criticality of cores present in the system. This is necessary when the shared memory sends data responses to requesting cores. For example, if a core has a read request to a cache line, and no other cores have a modified version of the cache line, the shared memory is responsible for supplying the data.

*2) Support for Timers:* Timer support in HourGlass also requires changes to the coherence protocol as timers allow cores to hold cache lines for a duration of time. We introduce new stable and transient coherence states to accommodate timer support in HourGlass. When a core receives data either from a remote core or from the shared memory, it moves to a stable state ($S$ or $M$ state), and populates both timer fields with timeout values, and begins to count down. When a core in a valid stable state such as shared ($S$) or modified ($M$) observes a request from a remote ncr- or crcore, it moves to a stable state $S^T I$ or $M^T I$, respectively. These states indicate that pending remote requests are observed by the core with a valid copy of the cache line, but the responses to these remote requests are deferred until timers timeout. Once the timers expire, the core issues a $SelfInv$ or $SendData$ coherence message to indicate that the core is either ready to self-invalidate its valid copy of the data or send the data to the requesting core. As a performance optimization, we add support for restarting timers if a core with a valid copy of the cache line does not observe remote write requests. These coherence messages contain the remote core identifier and its criticality information. The PRSP buffer introduced in Section VI-A holds these coherence messages and their information to allow the shared bus for canceling coherence responses from ncr cores in the presence of responses from cr cores. As mentioned in Section VI-A, these messages are issued by the bus arbitration in the TDM slot of the requesting core, which is available in the coherence message.

To avoid breaking the illusion of the single writer multiple reader (SWMR) invariant, which is an important property of cache coherence protocols, we introduce the coherence message $AllInv$ that is generated by the shared memory once all sharers have self-invalidated their copies of a cache line. Consider the following scenario where cores $c_0^{\mathsf{cr}}$ and $c_2^{\mathsf{ncr}}$ have valid copies of a cache line in shared ($S$) state, and $c_1^{\mathsf{cr}}$ has a pending write request to the same cache line. The pending write request needs to be satisfied after *all* sharers have self-invalidated their copies of the cache line. If the pending write request is satisfied otherwise, the system will result in at least two copies of the cache lines existing in the shared ($S$) state and modified ($M$) state simultaneously breaking the SWMR requirement. Hence, $SelfInv$ messages sent by the sharers are observed by the shared memory, and the sharer information in the shared memory is updated accordingly. Once all the sharers of a cache line have self-invalidated, the shared memory issues the $AllInv$ coherence message, which indicates updates to the cache line can proceed safely. Note that HourGlass currently does not allow for upgrades, which are optimized transitions from $S$ to $M$, and it does not allow a cache line in $M$ to move to $S$ state. For a cache line to move from $S$ to $M$ state on a write request, it has to wait for its own timer ($t_i(c_i^{\mathsf{cr}}, c_i^{\mathsf{cr}})$ or $t_i(c_i^{\mathsf{ncr}}, c_i^{\mathsf{ncr}})$) to timeout before proceeding with the write request. Section A-A presents an example that shows how HourGlass handles multiple pending requests from cr and ncr cores.

## VII. Timing Analysis

We derive the worst-case (WC) latency bound of a memory request, $L_i$ that a critical core, $c_i^{\mathsf{cr}}$, suffers due to interference from other cores upon accessing the shared cache hierarchy. $L_i$ has three latency components: arbitration (Definition 1), basic access (Definition 2), and coherence (Definition 3).

**Definition 1.** ***Arbitration latency***, *$L_{i,r}^{arb}$, of a request $r$ generated by $c_i^{\mathsf{cr}}$ is measured from the time stamp of its issuance until it is granted access to the bus. $L_{i,r}^{arb}$ occurs due to prior requests from other cores scheduled before $c_i^{\mathsf{cr}}$.*

**Definition 2.** ***Access latency*** *is the time required to transfer the requested data by $c_i^{\mathsf{cr}}$ from the shared memory to the private cache of $c_i^{\mathsf{cr}}$. We assume that the data transfer takes a fixed latency, $L^{acc}$.*

**Definition 3.** ***Coherence latency***, *$L_{i,r}^{coh}$, of a request $r$ generated by $c_i^{\mathsf{cr}}$ is measured from the time stamp when $r$ is granted access to the bus until it starts its data transfer. $L_{i,r}^{coh}$ is due to the coherence protocol rules that ensure data correctness in the cache hierarchy.*

A request generated by $c_i^{\mathsf{cr}}$ to a cache line $A$ suffers a **coherence latency** in three situations:

1) If another core, $c_j$, (critical or non-critical) has $A$ in a modified state. Thus, $c_i^{\mathsf{cr}}$ has to wait until $c_j$'s timer for $A$ expires before it gains an access to it.
2) If $C$ critical cores requested $A$ to be modified before $c_i^{\mathsf{cr}}$ issued its request, where $C \geq 1$. In this case, $c_i^{\mathsf{cr}}$ has to wait for $C$ cores to perform its operations and wait for its timers to expire before $c_i^{\mathsf{cr}}$ gets access to $A$.
3) If $c_i^{\mathsf{cr}}$ has $A$ in shared state and wants to write to it. In this case, $c_i^{\mathsf{cr}}$ must wait for $A$'s timer to expire in its private cache before $c_i^{\mathsf{cr}}$ issues its own write request to $A$.

**Lemma 1.** *The WC arbitration latency of any request generated by a critical core $c_i^{\mathsf{cr}}$ can be calculated as:*

$$WCL_i^{arb} = N_{\mathsf{cr}} \times SW$$

**Lemma 2.** *The WC coherence latency, $WCL_i^{coh}$, for a request issued by $c_i^{\mathsf{cr}}$ to a cache line A occurs under the following critical instance.*

*1) $c_i^{\mathsf{cr}}$ issues a write request to A, which $c_i^{\mathsf{cr}}$ has just received to be in the shared state in its private cache.*
*2) Afterwards, before the expiration of A's timer in $c_i^{\mathsf{cr}}$'s private cache, a non-critical core $c_j^{\mathsf{ncr}}$ obtains A in the shared state.*
*3) Before $c_j^{\mathsf{ncr}}$ times out, another critical core $c_k^{\mathsf{cr}}$ requests A in the modified state. Hence, $c_k^{\mathsf{cr}}$ has to wait for $v(ncr, cr)$ for $c_j^{\mathsf{ncr}}$ to time out. In addition, the self-invalidation message of $c_j^{\mathsf{ncr}}$ will be issued in $c_k^{\mathsf{cr}}$'s slot, which in the worst-case will have to wait for $N_{cr} - 1$ slots.*
*4) Finally, in the period between action 2) and the expiration of A's timer in $c_i^{\mathsf{cr}}$'s cache, all other ($N_{\mathsf{cr}} - 1$) critical cores issue write requests to A.*

*Accordingly, its WC coherence latency is:*

$$\begin{aligned} WCL_i^{coh} &= v(\mathsf{cr}, \mathsf{cr}) \\ &+ \big(v(\mathsf{ncr}, \mathsf{cr}) + (N_{\mathsf{cr}} - 1) \times SW\big) \\ &+ \big((N_{\mathsf{cr}} - 1) \times (v(\mathsf{cr}, \mathsf{cr}) + (N_{\mathsf{cr}} - 1) \times SW)\big) \\ &- N_{\mathsf{cr}} \times SW \end{aligned} \tag{1}$$

We present the proof for this lemma in Section A-B.

**Theorem 1.** *From Lemmas 1 and 2, and given that the access to the cache takes one TDM slot of $SW$ cycles, the total WC latency of a request by $c_i^{\mathsf{cr}}$ is as follows.*

$$WCL_i^{tot} = WCL_i^{arb} + WCL_i^{coh} + L^{acc} \tag{2}$$

*Proof.* The total WCL is the summation of the arbitration, access, and coherence latencies. □

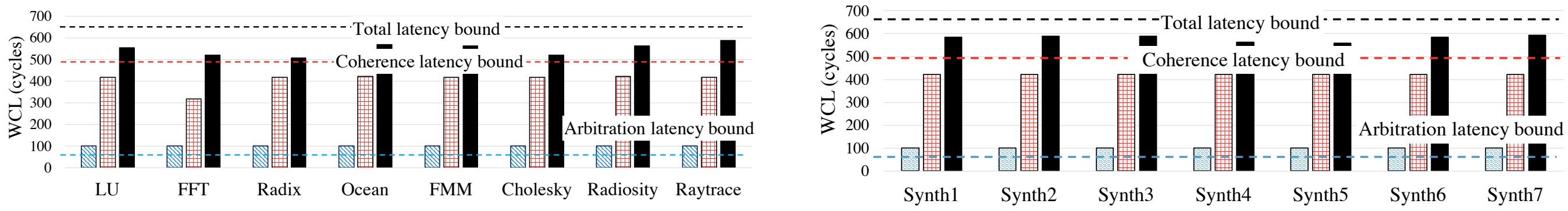

(a) SPLASH-2 benchmark suite.

(b) Synthetic benchmarks.

Fig. 3: Observed worst-case latency components for cr cores.

## VIII. Evaluation

We implement HourGlass[1] in gem5 [4], which is a cycle accurate micro-architectural simulator. We use the Ruby memory model to implement the cache, memory subsystem, and coherence protocol with high precision. Our multi-core setup consists of four x86 in-order cores that run at 2GHz. We use two cr cores $c_0^{\mathsf{cr}}, c_1^{\mathsf{cr}}$, and two ncr cores $c_2^{\mathsf{ncr}}, c_3^{\mathsf{ncr}}$. Every core has a separate 16kB direct mapped private L1-data (L1-D) and instruction (L1-I) cache with a cache line size of 64B. We use a 3 cycle access latency to the private L1 caches. All the cores share a 8-way 1MB set-associative last-level cache (LLC). Accesses to the LLC are assumed to be perfect and incur an access latency of 50 cycles. This allows our evaluation to focus on measuring metrics relevant to cache coherence. We assume that main-memory access overheads can be calculated using prior approaches such as [19], and are additive to the coherence latencies derived in this work [20]. The bus manages accesses between cores to the shared memory using the policy described earlier. For our evaluation, we use SPLASH-2 [5] as a multi-threaded benchmark suite, and multi-threaded synthetic benchmarks for maximum sharing of data across cores. We run these benchmarks using four threads, and we map each thread to a core.

### A. Data and Protocol Verification

We verify the correctness of HourGlass using synthetic and SPLASH-2 benchmarks. The synthetic benchmarks exercise maximum sharing of data by executing the same set of instructions across all cores on the same data. With these, we manually verify that all state transitions, and timer events of HourGlass are correctly exercised. This confirms correctness of HourGlass. For data correctness of HourGlass, we check the output of SPLASH-2 benchmark suite. These benchmarks have in-built single threaded verification routines that check the output generated from the multi-threaded workload.

### B. Bounding memory access latencies

We show that with HourGlass, cr cores have a bound on the access latency to shared data. Figure 3 shows the observed worst-case request access latency, and the individual latency components such as the coherence latency, and arbitration latency for the cr cores using SPLASH-2 and synthetic benchmarks. In this evaluation, we configure the initial timer values as follows: $v(\mathsf{cr},\mathsf{cr})$=$2 * TDM$, $v(\mathsf{cr},\mathsf{ncr})$=$4 * TDM$, $v(\mathsf{ncr},\mathsf{cr})$=$1 * TDM$, and $v(\mathsf{ncr},\mathsf{ncr})$=$2 * TDM$.

We observe in Figure 3 that the WCL for all benchmarks, and their respective components are within the analytical bounds. We note that SPLASH-2 is designed such that there is minimal data sharing across multiple threads. Hence, the observed WCLs in Figure 3a for the different components are also well within the analytical bounds. Figure 3b shows that WCL of cr cores using synthetic benchmarks that perform the same set of instructions all cores. We observe that for synthetic workloads that stress the coherence protocol, the WCL are within the analytical bounds.

### C. Comparison with PMSI

Figure 4 shows the WCL of a memory request from cr cores and ncr cores deployed using HourGlass and PMSI. The PMSI protocol is modified to support cache-to-cache transfers, and hence we compute the analytical WCL for a memory request under this variant of PMSI accordingly. We also compare with a HourGlass configuration that sets the initial timer values to zero (HourGlass (0,0,0,0)). For this evaluation, we use one of the synthetic benchmarks (Synth1) that stresses the worst-case scenarios for both protocols.

When the timer values are zero, HourGlass provides tighter bounds on the WCL of cr cores compared to that of PMSI. This is because the criticality-aware bus arbitration policy guarantees slots only for cr cores resulting in lower arbitration latencies. Moreover, by setting the timer values to zero, the WCL of cr cores is further reduced. The mixed timer configuration of HourGlass has the same WCL bound for cr cores compared to PMSI. This is because the timer configurations affects the WCL of cr cores. On the other hand, PMSI does not differentiate between different critical cores, and hence the WCL of a memory request incurs additional arbitration latency for arbitrating across ncr cores. Another consequence of the criticality aware arbitration is the increase in WCL of ncr cores in HourGlass compared to PMSI. WCL of ncr cores with PMSI have the same WCL of cr cores as it does not differentiate criticality of memory requests. HourGlass on the other hand grants slots to ncr cores in slack slots of cr cores, which is dependent on the application.

[1] https://git.uwaterloo.ca/caesr-pub/hourglass

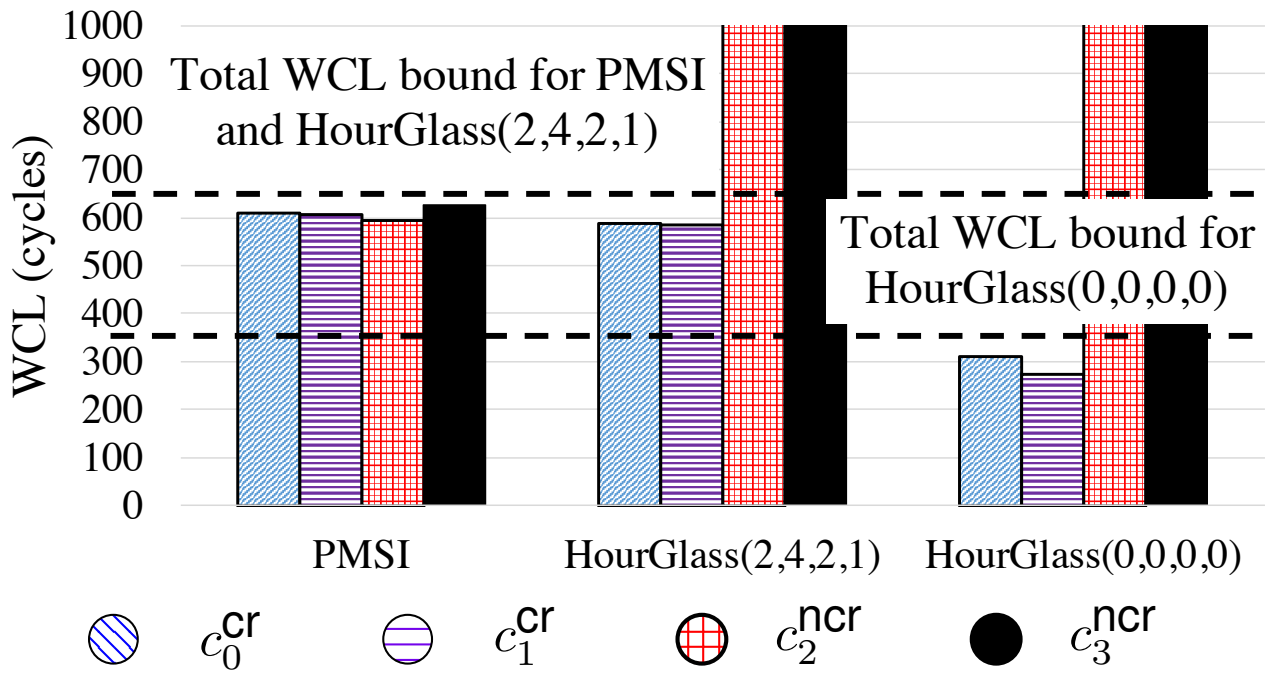

Fig. 4: WCL of cr cores using HourGlass and PMSI.

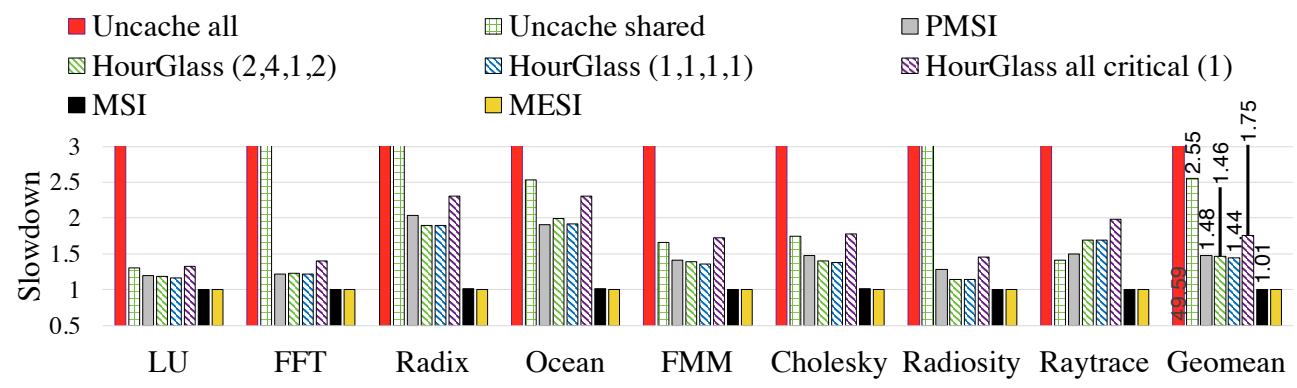

Fig. 5: Execution time slowdown compared to MESI protocol.

| Configuration | Timer values | Description |
|---|---|---|
| Hourglass(2,4,1,2) | $v(\text{cr}, \text{cr}) = 2 * P$, $v(\text{cr}, \text{ncr}) = 4 * P$, $v(\text{ncr}, \text{cr}) = 1 * P$, $v(\text{ncr}, \text{ncr}) = 2 * P$ | Different timer values based on criticality of cores that have valid copy and requesting cores |
| Hourglass(1,1,1,1) | $v(\text{cr}, \text{cr}) = 1 * P$, $v(\text{cr}, \text{ncr}) = 1 * P$, $v(\text{ncr}, \text{cr}) = 1 * P$, $v(\text{ncr}, \text{ncr}) = 1 * P$ | Same timer values irrespective of criticality of cores that have valid copy and requesting cores |
| Hourglass(1) | $v(\text{cr}, \text{cr}) = 1 * P$ | All cores are cr cores |

TABLE II: HourGlass configurations.

### D. Comparison with real-time approaches for managing shared data and conventional cache coherence protocols

We compare the performance of HourGlass with conventional cache coherence protocols such as the MSI and MESI protocols, state-of-the-art real-time cache coherence protocol PMSI, and other real-time approaches for sharing data predictably in Figure 5. These approaches include *uncache-all* that does not cache any data in the private caches, and *uncache-shared* [8] that caches only private data and accesses shared data from shared memory. Since, HourGlass is the only cache coherence protocol for DCMS, we present three different HourGlass configurations for comparison. We tabulate the configurations in Table II and their descriptions. The purpose of having the HourGlass (1,1,1,1) configuration is to observe the effect of slowdown suffered by HourGlass when all the timer values are uniform but still be criticality-aware. Such a configuration may have varying slowdowns compared to the default HourGlass (2,4,1,2) configuration based on the application. We also evaluate the HourGlass (1) configuration that does not have criticality-awareness, similar to PMSI. However, cache lines are held in the private caches of the cores for a duration of 1 TDM period. We use the SPLASH-2 benchmark suite for this evaluation, and compute the slowdown experienced by the different protocols and approaches compared to the MESI cache coherence protocol.

*uncache-all* has the largest slowdown compared to the rest of the approaches and protocols. On average it suffers from a 49.59× slowdown as the SPLASH-2 benchmarks exhibit data reuse. This is because every data access incurs the long memory access latency. *uncache-shared* performs significantly better than *uncache-all* as some of the private data reuse are cache hits in the private caches. On average, *uncache-shared* suffers a slowdown of 2.55× compared to the MESI protocol. PMSI protocol outperforms previous real-time approaches as it allows for multiple copies of shared and private data to be cached in the private caches. Hence, PMSI experiences less average slowdown (48%) compared to the previous approaches. We observe that HourGlass with criticality-awareness (HourGlass (2,4,1,2) and HourGlass (1,1,1,1) configurations performs slightly better compared to PMSI protocol. In particular, HourGlass (1,1,1,1) improves execution time by 2% on average over PMSI. We observe that almost all SPLASH-2 benchmarks partition the data across cores such that there is minimal sharing, and have high data reuse resulting in high cache hit rates. Hence, we observe that there are abundant slack slots for ncr cores to satisfy their memory requests. Moreover, the SPLASH-2 benchmarks include synchronization of threads in the form of barriers that result in cr cores waiting for ncr cores to satisfy the synchronization condition. This combination results in comparable execution times between HourGlass and PMSI for the SPLASH-2 benchmark suite. The HourGlass (1) configuration that considers all cores to be cr cores also performs better than the alternative real-time approaches. However, compared to previous HourGlass configurations and PMSI , this configuration experiences more slowdown compared to the MESI protocol. In particular, HourGlass (1) incurs on average a 20% and 19% increase in execution time compared to the HourGlass (2,4,1,2) configuration and PMSI, and experiences an average slowdown of 75% compared to the MESI protocol. Although SPLASH-2 benchmarks exhibit data reuse and locality, the data reuse do not occur within the timer duration. Hence, cores hold on to cache lines within a duration where they may not be reused, and unnecessarily stall cores requesting for the same cache line. This is exacerbated with calls to synchronization routines in SPLASH-2 benchmarks resulting in increased slowdown for this configuration.

### E. Effect of timers on bandwidth utilization of ncr cores

For DCMS, timing guarantees are necessary only for cr cores. ncr cores on the other hand do not have timing guarantees but it may be desirable to improve their memory bandwidth utilization. This is because applications running on ncr cores may have high locality and hence high execution throughput. We observe that with HourGlass, varying memory

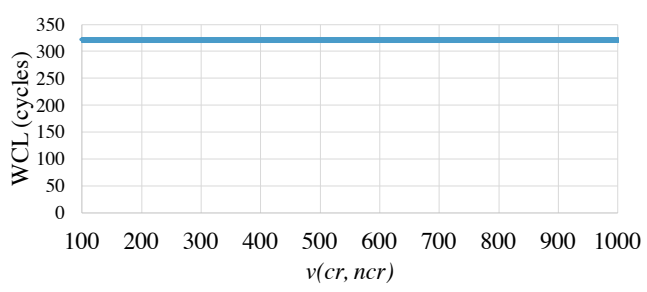

(a) WCL of cr cores.

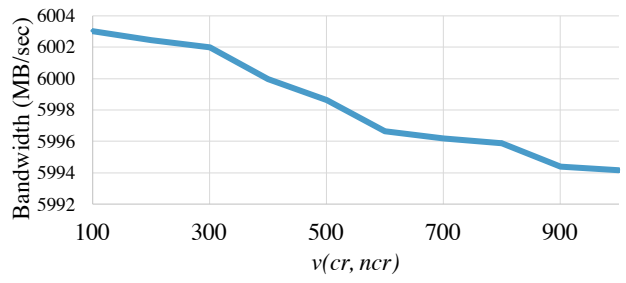

(b) Memory bandwidth utilization of ncr cores.

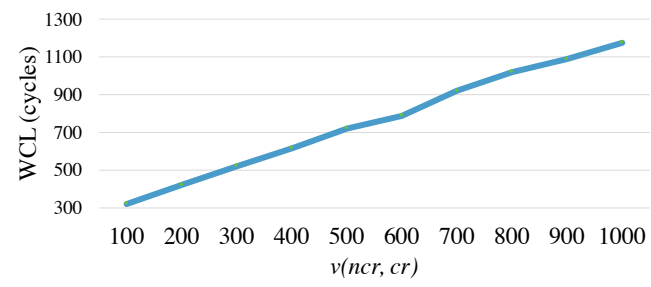

(c) WCL of cr cores.

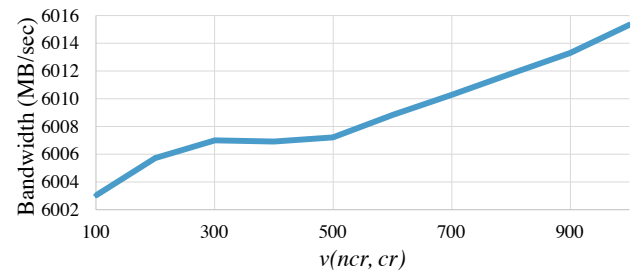

(d) Memory bandwidth utilization of ncr cores.

Fig. 6: Effect on WCL and memory bandwidth with varying $v(\mathsf{cr}, \mathsf{ncr})$ and $v(\mathsf{ncr}, \mathsf{cr})$.

bandwidth utilizations can be achieved for ncr cores at the expense of increasing the WCL bound for cr cores. This is because the time duration for which ncr cores core holds data in its private cache ($v(\mathsf{ncr}, \mathsf{cr})$) affects the WCL of the cr cores. We show the effects of varying two initial timer values $v(\mathsf{cr}, \mathsf{ncr})$ and $v(\mathsf{ncr}, \mathsf{cr})$ on the memory bandwidth utilization of ncr cores and WCL of cr cores in Figure 6. We use the Cholesky benchmark from SPLASH-2 as it has the lowest cache hit ratio across all the benchmarks.

From Figure 6a, the WCL of cr cores does not change with varying $v(\mathsf{cr}, \mathsf{ncr})$. However, we observe a decrease in memory bandwidth for ncr cores with increase in $v(\mathsf{cr}, \mathsf{ncr})$ as shown in Figure 6b. This is because $v(\mathsf{cr}, \mathsf{ncr})$ affects the time duration for which a cr cores core holds data in its private cache in the presence of pending requests from ncr cores. A high value of $v(\mathsf{cr}, \mathsf{ncr})$ defers responses to pending requests from ncr cores resulting in more opportunities for observing responses from cr cores and hence, reordering responses. Varying $v(\mathsf{ncr}, \mathsf{cr})$ on the other hand affect the WCL of cr cores, and memory bandwidth utilization of ncr cores as shown in Figures 6c and 6d respectively. In Figure 6c, higher values of $v(\mathsf{ncr}, \mathsf{cr})$ increases the WCL of cr cores. This is because a request from cr cores core to a shared data that is in the private cache of ncr cores core has to wait for a maximum of $v(\mathsf{ncr}, \mathsf{cr})$ before completing its request. However, increasing $v(\mathsf{ncr}, \mathsf{cr})$ allows ncr cores to increase cache hits on shared data cached in their private caches resulting in improved memory bandwidth. This is shown in Figure 6d. Hence, the timers in HourGlass provides both bounds on WCL for cr cores, and satisfy memory bandwidth requirements for ncr cores at the expense of increasing the WCL of cr cores.

## IX. Conclusion

In this work, we present HourGlass, a predictable time-based cache coherence protocol that provides WCL bounds for cr cores in a DCMS. HourGlass also includes provisions to improve the memory bandwidth utilization for ncr cores at the expense of increasing the WCL of cr cores. Since HourGlass is criticality-aware, it provides tighter WCL bounds for cr cores compared to the state-of-the-art real-time cache coherence protocol. HourGlass exhibits similar total execution times compared to the state-of-the-art real-time cache coherence protocol, and experiences a 1.43× and 1.46× slowdown compared to the conventional unpredictable MSI and MESI protocols for SPLASH-2 benchmarks.

## Appendix A

### A. Example scenario with HourGlass

Figure 7 shows a scenario with multiple pending requests to a shared cache line $A$ using HourGlass. The arbitration policy is criticality aware, and we presented the arbitration details for the criticality-aware arbitration policy in Section IV. Initially all cores do not have a valid copy of the cache line $A$. The initial timeout value for all timer values is 2 TDM periods.

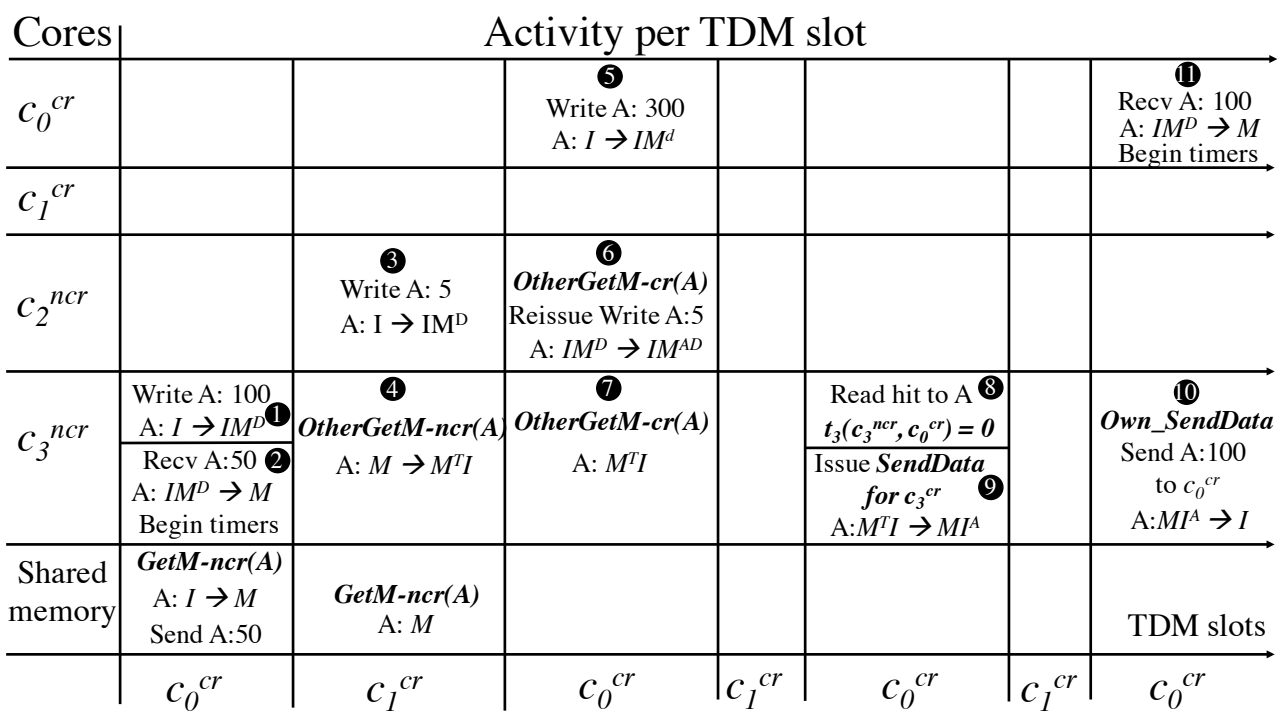

Fig. 7: Example scenario with HourGlass.

At ❶, $c_3^{\mathsf{ncr}}$ has a pending write request to $A$, which is serviced in the slack slot of $c_0^{\mathsf{cr}}$. Since, there are no sharers for this cache line, the shared memory responds with the data, and changes its state to $M$ ❷. On receiving the data, the two timers for the cache lines are loaded with the initial timeout values and begin to count down. Since $c_1^{\mathsf{cr}}$ does not have a pending request, and $c_2^{\mathsf{ncr}}$ has a pending write request to $A$, the bus arbitration slot grants this slack slot to $c_2^{\mathsf{ncr}}$ for its write request ❸. At ❹, $c_3^{\mathsf{ncr}}$ notices the write request from $c_2^{\mathsf{ncr}}$ on the bus, which is denoted as $OtherGetM$-ncr, and changes its state from $M$ to $M^T I$. It also marks the ncr-destination field in its modified copy of $A$ as $c_2^{\mathsf{ncr}}$. In the next TDM slot of $c_0^{\mathsf{cr}}$, $c_0^{\mathsf{cr}}$ has a pending write request to $A$ that is serviced by the shared bus ❺. At ❻, $c_2^{ncr}$ observes cr core $c_0^{cr}$'s write request, and reissues its write request. At the same time ❼, $c_3^{\mathsf{ncr}}$ observes this write request, and updates the cr-destination field as $c_0^{\mathsf{cr}}$. Since, the state of $A$ in $c_3^{\mathsf{ncr}}$ is $M^T I$, and is waiting for its timer to expire, it does not respond to the request from $c_0^{\mathsf{cr}}$ yet. At ❽, a read request from $c_3^{\mathsf{ncr}}$ to cache line $A$ is a cache hit. Notice that this is a cache hit because of the timer support, thereby improving the memory bandwidth utilization. In the absence of timers, the previous pending write requests to the same cache line would have resulted in $A$'s invalidation in $c_3^{\mathsf{ncr}}$ resulting in a cache miss for this read request. At ❾, the timer $t_3(c_3^{\mathsf{ncr}}, c_0^{\mathsf{cr}})$ counts down to 0, and issues a $SendData$ message with the requester core information set to that of the cr core $c_0^{\mathsf{cr}}$ [2]. $c_3^{\mathsf{ncr}}$ moves from $M^T I$ to $M^I A$ state, which denotes that it is waiting for the $SendData$ message to be ordered on the shared bus. The data transfer between $c_3^{\mathsf{ncr}}$ and $c_0^{\mathsf{cr}}$ is done in the requesting core's slot, which is $c_0^{\mathsf{cr}}$ as ncr cores are not guaranteed slots by the arbiter. At ❿, $c_3^{\mathsf{ncr}}$'s $SendData$ message is observed on the bus, and $c_3^{\mathsf{ncr}}$ transfers the up-to-date $A$ data to $c_0^{\mathsf{cr}}$. At ⓫, $c_0^{\mathsf{cr}}$ performs the write operation and moves from $IM^D$ to $M$ state.

[2]This occurs if $v(\mathsf{ncr}, \mathsf{cr}) < v(\mathsf{ncr}, \mathsf{ncr})$. If $v(\mathsf{ncr}, \mathsf{ncr}) > v(\mathsf{ncr}, \mathsf{cr})$, the cache controller will lookup the destination fields for the cache line and issue the $SendData$ message based on the whether a cr core is pending or not.

### B. Proof for Lemma 2

*Proof.* This proof is by contradiction.

1) Let the WC for $c_i^{\mathsf{cr}}$ be such that $c_i^{\mathsf{cr}}$ does not have $A$ in its private cache. Then, $c_i^{\mathsf{cr}}$ does not have to wait until its timer expires before it issues its own write request to $A$. This means $c_i^{\mathsf{cr}}$ no longer waits for $v(\mathsf{cr}, \mathsf{cr})$ (first term in equation 1). Therefore, the resulting latency is less than $WCL_i^{coh}$. Thus, it cannot be the WC.
2) Let the WC be when $A$ is not cached in the private cache of any non-critical core upon the expiration of $A$'s timer of $c_i^{\mathsf{cr}}$. In this case, $c_i^{\mathsf{cr}}$ no longer waits; hence, the non-critical core does not issue any coherence messages $(\big(v(\mathsf{ncr}, \mathsf{cr}) + (N_{\mathsf{cr}} - 1) \times SW\big)$ is removed.) Therefore, the resulting latency is less than $WCL_i^{coh}$.
3) Let the WC be when $N'$ critical cores issue write requests to $A$ before $A$'s timer expiration of $c_i^{\mathsf{cr}}$ such that $N' < N_{\mathsf{cr}} - 1$ where the remaining $N_{\mathsf{cr}} - N' - 1$ cores have either no request or are requesting access to a different cache line $B$. Therefore, $c_i^{\mathsf{cr}}$ waits for each $N'$ core to access the data, and $v(cr, cr)$ before $c_i^{\mathsf{cr}}$ gets access to $A$. Since $N' < N_{\mathsf{cr}} - 1$, $N' \times \big(v(\mathsf{cr}, \mathsf{cr}) + (N_{\mathsf{cr}} - 1) \times SW\big)$ is less than the value of the third term in Equation 1, and hence it cannot be the WC.

Finally, for all the other critical cores to request $A$ before the expiration of $c_i^{\mathsf{cr}}$'s timer, they need at least one TDM period. This justifies the subtracted term in Equation 1. □

### C. Scalability

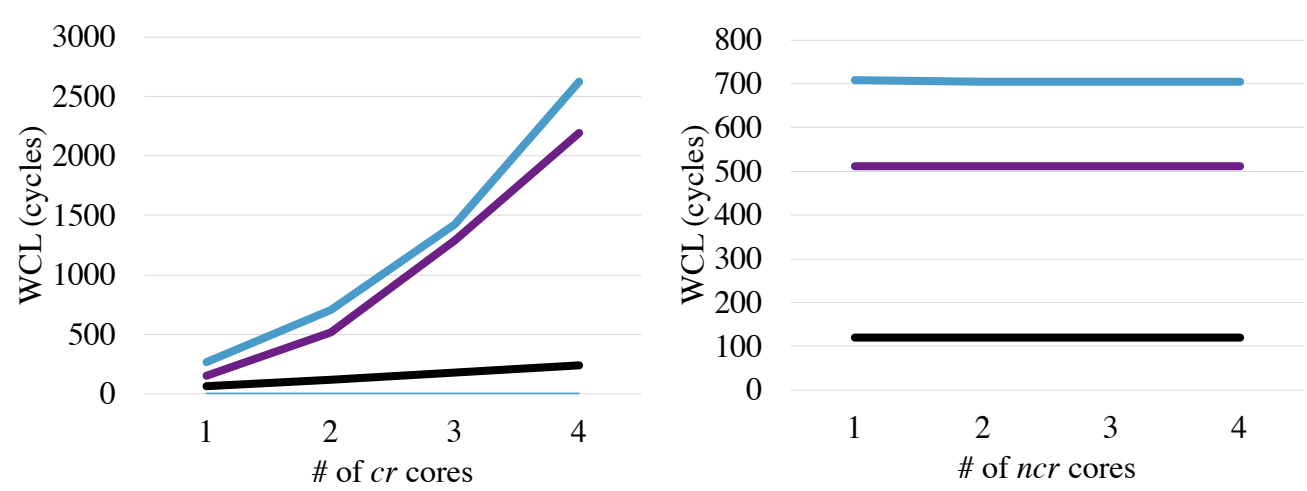

(a) Scaling cr cores. (b) Scaling ncr cores.

Total latency  Coherence latency  Arbitration latency

Fig. 8: Effect of scalability on WCL of cr cores.

We analyzed the effects of varying the number of cr- and ncr cores on the WCL of cr cores. Recall that the WCL of cr cores is dependent on the number of cr cores, and independent of the number of ncr cores. Hence, varying the number of ncr cores should not affect the WCL of cr cores. We show these trends using Figure 8 where we vary the number of cr cores and ncr cores respectively. In Figure 8a, we vary the number of cr cores and keep the number of ncr cores to a constant, and in Figure 8b, we vary the number of ncr cores and keep the number of cr cores constant. For this evaluation, we use one of the synthetic benchmarks to distinctly highlight the trends as the synthetic benchmarks have maximum sharing of data across the cores[3]. In Figure 8a, the WCL of cr cores increases with increase in the number of cr cores. In particular, the arbitration latency is proportional to the TDM period, which is a function of number of cr cores, and the coherence latency depends on the timer values, which in turn are fixed in terms of TDM periods. In addition, the coherence latency increases quadratically with increase in the number of cr cores, and dominates the total coherence latency. However, from Figure 8b, the WCL of cr cores does not increase with the number of ncr cores. However, increasing the number of ncr cores may affect memory bandwidth utilization for ncr cores, as they contend for slack slots of the cr cores.

| State | Core events | | | | Bus events - common to cr and ncr cores | | | | | | | | |
|---|---|---|---|---|---|---|---|---|---|---|---|---|---|
| | Load | Store | Replacement | Timeout | OwnGetS | OwnGetM | OwnPutM | OwnSelfInv | InvAll | OwnSendData | Data | OtherGetS-cr or ncr | OtherGetM-cr or ncr |
| I | issue GetS/IS$^{AD}$ | issue GetM/IM$^{AD}$ | X | X | X | X | X | X | X | X | X | - | - |
| IS$^{AD}$ | X | X | X | X | /IS$^{D}$ | X | X | X | X | X | X | - | - |
| IS$^{D}$ | X | X | X | X | X | X | X | X | X | X | load/S and ST | * | * |
| IS$^{D}$I | X | X | X | X | X | X | X | X | X | X | load/S$^{T}$I and ST | * | * |
| S | hit | /S$^{T}$M and WT | issue SelfInv/SI$^{A}$ | /S and RT | X | X | X | X | X | X | X | - | update (cr or ncr)-Dest/ S$^{T}$I and WT |
| S$^{T}$I | hit | S$^{T}$M and WT | issue SelfInv/SI$^{A}$ | issue SelfInv/SI$^{A}$ | X | X | X | X | X | X | X | - | update (cr or ncr)-Dest |
| SI$^{A}$ | hit | issue SelfInv and GetM/SM$^{A}$ | issue SelfInv | X | X | X | X | /SI | X | X | X | - | issue SelfInv |
| SI | hit | issue GetM/IM$^{AD}$ | /I | X | X | X | X | X | /I | X | X | - | - |
| S$^{T}$M | hit | X | X | issue SelfInv and GetM/SM$^{A}$ | X | X | X | X | X | X | X | - | update (cr or ncr)-Dest |
| SM$^{A}$ | hit | X | X | X | X | X | X | /IM$^{AD}$ | X | X | X | - | issue SelfInv |
| IM$^{AD}$ | X | X | X | X | X | /IM$^{D}$ | X | X | X | X | X | - | - |
| IM$^{D}$ | X | X | X | X | X | X | X | X | X | X | store/M and ST | * | * |
| IM$^{D}$I | X | X | X | X | X | X | X | X | X | X | store/M$^{T}$I and ST | * | * |
| M | hit | hit | issue PutM/MI$^{R}$ | /M and RT | X | X | X | X | X | X | X | /M$^{T}$I and WT | update (cr or ncr)-Dest / M$^{T}$I and WT |
| M$^{T}$I | hit | hit | issue PutM and SendData/MI$^{R}$ | issue SendData/MI$^{A}$ | X | X | X | X | X | X | X | update (cr or ncr)-Dest | update (cr or ncr)-Dest |
| MI$^{R}$ | hit | hit | X | X | X | X | WB/I | X | X | issue SendData/I | X | issue SendData | issue SendData |
| MI$^{A}$ | hit | hit | issue PutM/MI$^{R}$ | X | X | X | WB/I | X | X | issue SendData/I | X | issue SendData | issue SendData |

TABLE A1: HourGlass private cache coherence states. Shaded rows are new states introduced by HourGlass. WT: Wait for timer timeout, RT: Restart timer, ST: Start timer. $msg/state$ denotes that a core issues the message $msg$, and moves to coherence state $state$. Cells marked as "$-$" indicate that a particular transition cannot happen, and cells marked as "$\times$" denote that a cache line in that state does not change state with a core event or bus event.

### D. Hardware overhead of HourGlass

We briefly summarize the hardware overhead for HourGlass on a 4-core system. Each cache line in the private caches are extended by 2 timer fields, and two destination fields resulting in an overhead of $64 + 64 + log_2(4) + log_2(4) = 132$ bits. Each cache line in the shared memory includes the sharer bits, which for a 4-core system results in an overhead of 4 bits per cache line. Each entry in the PR LUT in the shared memory is extended by a bit to denote the criticality of the request in a DCMS. Each entry in the PRSP buffer holds the memory address for the request, the coherence message, the core identifier, and a valid bit. For a physical address space of 4GB, this results in roughly 40 bits per entry.

[3]Note that the other synthetic benchmarks show the same trends, and hence we show only one of the synthetic benchmarks. We do not show the evaluation of this experiment on SPLASH-2 benchmarks as they have minimal data sharing, and therefore, the trend lines are not as distinct as the synthetic benchmarks.